\documentclass[submission, Proceedings]{SciPost}

\begin{document}

\begin{center}{\Large \textbf{
$\mu$-$e$ conversion experiments at J-PARC
}}\end{center}

% TODO: write the author list here. Use initials + surname format.
% Separate subsequent authors by a comma, omit comma at the end of the list.
% Mark the corresponding author with a superscript *.
\begin{center}
N. Teshima\textsuperscript{1*}
\end{center}

\begin{center}
{\bf 1} Osaka City University, Osaka, Japan
\\
* teshima@ocupc1.hep.osaka-cu.ac.jp
\end{center}

\begin{center}
\today
%November 15, 2018
\end{center}

\definecolor{palegray}{gray}{0.95}
\begin{center}
\colorbox{palegray}{
  \begin{tabular}{rr}
    \begin{minipage}{0.05\textwidth}
      \includegraphics[width=8mm]{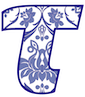}
    \end{minipage}
    &
    \begin{minipage}{0.82\textwidth}
      \begin{center}
        {\it Proceedings for the 15th International Workshop on Tau Lepton Physics,}\\
        {\it Amsterdam, The Netherlands, 24-28 September 2018} \\
        \href{https://scipost.org/SciPostPhysProc.1}{\small \sf scipost.org/SciPostPhysProc.Tau2018}\\
      \end{center}
    \end{minipage}
  \end{tabular}
}
\end{center}

% For convenience during refereeing: line numbers
%\linenumbers

\section*{Abstract}
{\bf
% TODO: write your abstract here.
There are two plans of experiments to search for muon to electron ($\mu$-$e$) conversion at J-PARC, which are called DeeMe and COMET. $\mu$-$e$ conversion is one of the charged lepton flavor violation processes, which are forbidden in the Standard Model, but some theories beyond the Standard Model predict relatively large branching ratios at orders of $10^{-12}$ to $10^{-17}$. DeeMe will be conducted with a sensitivity of $1\times10^{-13}$ using a carbon target for 1 year, and COMET will be done with that of $3\times10^{-15}$ in Phase-I. In this article, the current status of these two experiments will be presented.
%The abstract is in boldface, and should fit in 8 lines.
%It should be written in a clear and accessible style, emphasizing the context, the problem(s) studied, the methods used, the results obtained, the conclusions reached, and the outlook. You can add a table contents, recommended if your paper is more than 6 pages long.
}

\vspace{10pt}
\noindent\rule{\textwidth}{1pt}
\tableofcontents\thispagestyle{fancy}
\noindent\rule{\textwidth}{1pt}
\vspace{10pt}

\section{Introduction}
% TODO: write your article here.
%The stage is yours. Write your article here.
%The bulk of the paper should be clearly divided into sections with short descriptive titles, including an introduction and a conclusion.
\subsection{Reactions of Muonic Atoms and $\mu$-$e$ Conversion}
In the $\mu$-$e$ conversion search experiments, we stop negative muons in a target to form muonic atoms. The muon of 1S bound state in a muonic atom can decay, called the decay-in-orbit, or be captured by the nucleus through the charged-current weak interaction of the Standard Model. \\
\indent
The probabilities of those processes depend on the nuclear mass. For heavier nuclei, the probability of capture is higher, and the life time of muonic atom becomes shorter. For muonic carbon atoms, which the DeeMe experiment plans to use, 92\% of muons decay in orbit, while 8\% of them are captured by the nucleus. The life time of the muonic carbon atom is $2.0\ \mu\mathrm{s}$. The COMET experiment uses an aluminum target. The life time of the muonic aluminum is $0.86\ \mu\mathrm{s}$, where 39\% is the decay-in-orbit and 61\% is the muon capture. In the DeeMe experiment, a silicon-carbide target is being investigated as a target. A muonic silicon atom has a life time of $0.76\ \mu\mathrm{s}$ with the decay-in-orbit 33\% and the muon capture 66\%. \\
\indent
We search for $\mu$-$e$ conversion, which is one of the charged lepton flavor violating processes. It is the coherent neutrino-less conversion of a muon into an electron in the nuclear field. The conversion electron is monoenergetic with an energy equal to the muon mass minus the binding energy of the muon and the nuclear recoil. It will be approximately $105\ \mathrm{MeV}$ for the target atom made of carbon, aluminum or silicon.
\subsection{Sensitivity Goals}
The current limits on the branching ratio of the $\mu$-$e$ conversion are $4.6\times10^{-12}$ for a titanium target by an experiment at TRIUMF \cite{1}, $4.3\times10^{-12}$ for a titanium target \cite{2} and $7\times10^{-13}$ for a gold target \cite{3} by the SINDRUM-II experiment at PSI. \\
\indent
The goal of the DeeMe experiment is to achieve a single event sensitivity of $1\times10^{-13}$ for a graphite target for one year. If we change the target to silicon carbide, the sensitivity improves to be approximately $2\times 10^{-14}$. These sensitivities are better by one or two orders of magnitude than those achieved so far. \\
\indent
The goal of the COMET experiment is to achieve a sensitivity of $3\times10^{-15}$ in Phase-I and $2\times10^{-17}$ in Phase-II. These sensitivities are better by two to four orders of magnitude.
\subsection{Places of Experiments}
\begin{center}
  \begin{figure}[t]
    \begin{tabular}{ccc}
      \begin{minipage}[t]{0.28\hsize}
        \centering
        \includegraphics[width=4.2cm,keepaspectratio,clip]{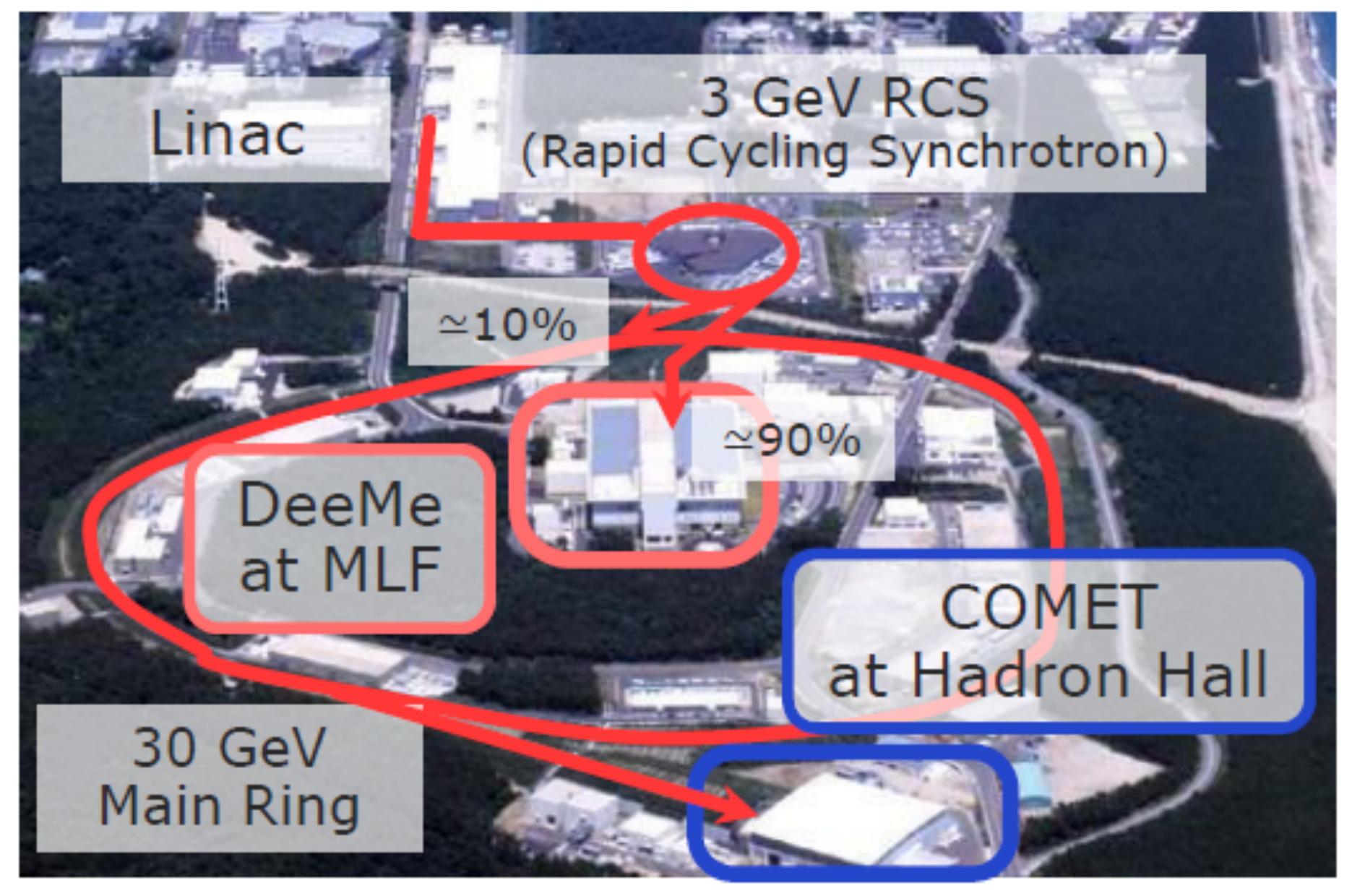}
        \caption{Places where the DeeMe and COMET experiments are conducted in J-PARC.}
        \label{fig:jparc}
      \end{minipage} &
      \begin{minipage}[t]{0.3\hsize}
        \centering
        \includegraphics[width=4.8cm,keepaspectratio,clip]{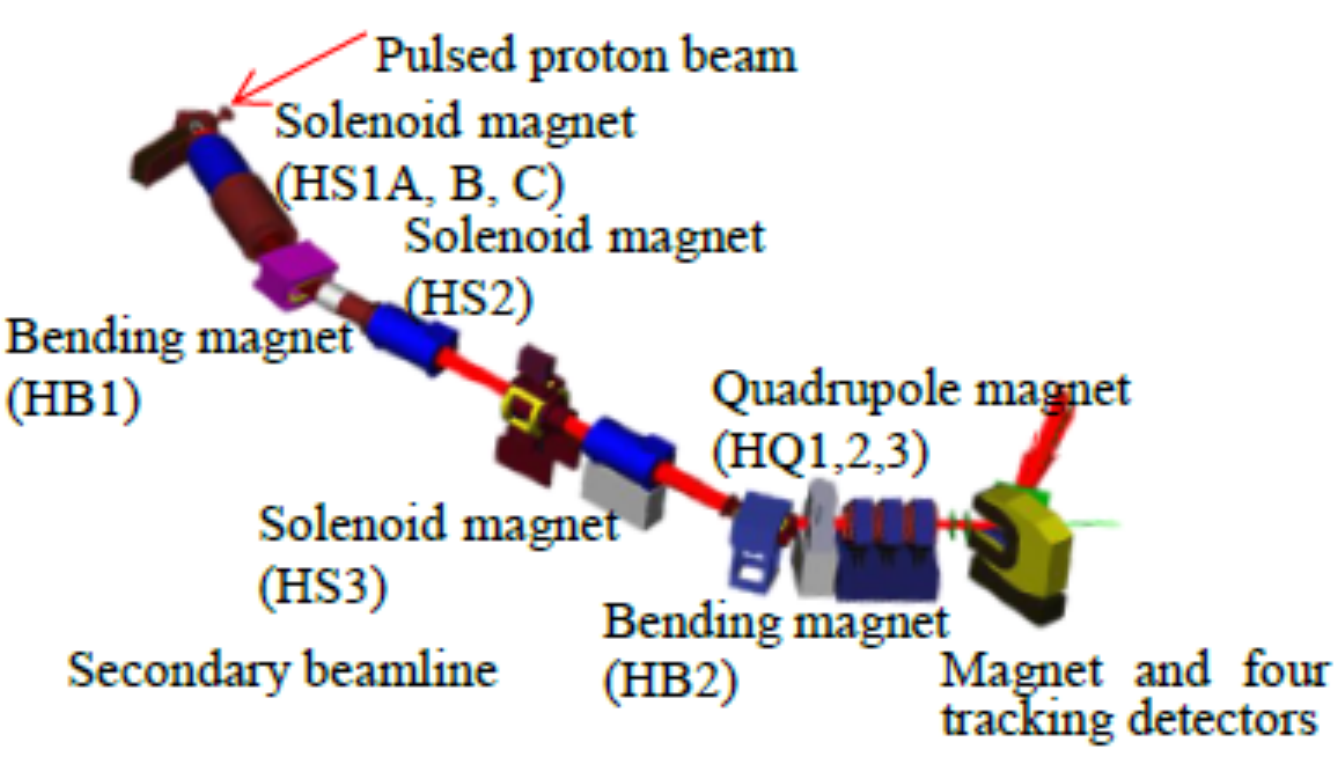}
        \caption{The experimental apparatus of the DeeMe experiment at J-PARC MLF H-Line.}
        \label{fig:hline}
      \end{minipage} &
      \begin{minipage}[t]{0.35\hsize}
        \centering
        \includegraphics[width=4.8cm,keepaspectratio,clip]{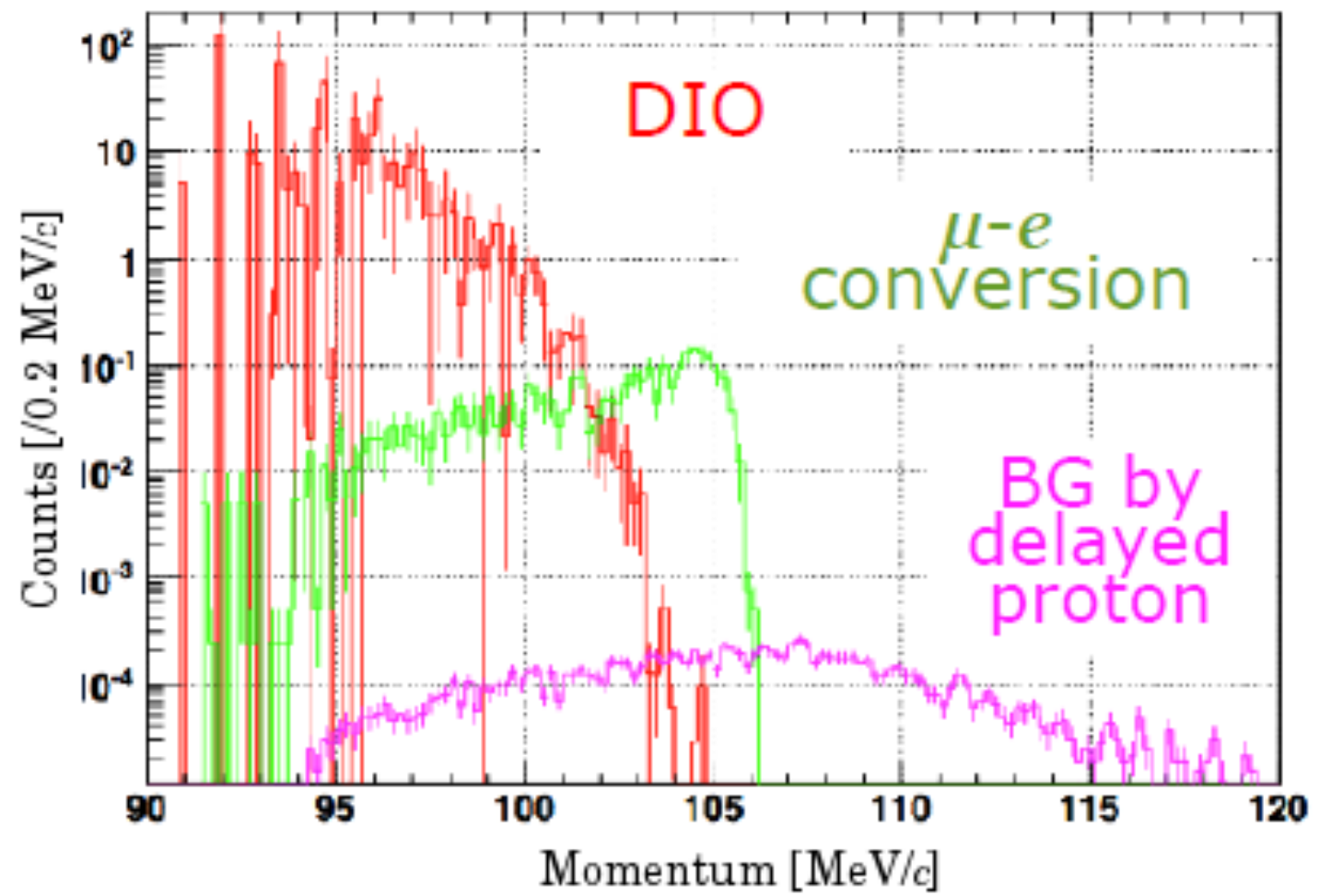}
        \caption{Simulated momentum spectra of electrons at the spectrometer.}
        \label{fig:backgrounddeeme}
      \end{minipage}      
    \end{tabular}
  \end{figure}
\end{center}

Two experiments are conducted at J-PARC in Tokai Village, Japan (see Fig. \ref{fig:jparc}). The place for the DeeMe experiment is Materials and Life Science Experimental Facility (MLF), and pulsed proton beam of fast extraction from 3-GeV Rapid Cycling Synchrotron (RCS) with a repetition of $25\ \mathrm{Hz}$ will be used. The COMET is located at Hadron Hall, and pulsed proton beam of slow extraction from the Main Ring will be used. The beam time does not conflict and these experiments can run in parallel.
\section{The DeeMe Experiment}
\label{sec:deeme}
The DeeMe Collaboration consists of thirty-seven people, representing thirteen institutes as of November, 2018.
\subsection{Experimental Concept}
Figure \ref{fig:hline} shows the experimental apparatus of the experiment. The pulsed proton beam is injected into the target to produce pions. The pions then decay in flight into muons, and muons are captured by the primary-target nuclei to form muonic atoms. \\
\indent
In typical experiments to search for $\mu$-$e$ conversion, a pion-production target, pion-decay and muon-transport section, and a muon-stopping target are separately prepared. But in the DeeMe experiment, only one primary-target will be used for all of the pion-production, pion-muon-decay, and muon-stopping. \\
\indent
The secondary beamline, H Line, transports $\mu$-$e$ conversion electrons from the primary target to a magnetic spectrometer. We will then search for the signal by using the spectrometer consisting of a dipole magnet and four tracking detectors, multi-wire proportional chambers (MWPCs).
\subsection{Backgrounds}
Figure \ref{fig:backgrounddeeme} shows simulated momentum spectra of electrons from muon decay-in-orbit, $\mu$-$e$ conversion, and beam-related background at the spectrometer for the RCS operating with $1\ \mathrm{MW}$ for $2\times10^{7}\ \mathrm{sec}$, using a silicon-carbide target, with the branching fraction of the $\mu$-$e$ conversion assumed to be $3\times10^{-14}$. \\
% delayed proton rate to the main pulse $< 10^{-19}$. \\
\indent
The low-momentum backgrounds will be suppressed by the secondary beamline, and a spectrometer with momentum resolution better than $1\ \mathrm{MeV/}c$ is sufficient to separate the signal from the high-momentum tail of the background. \\
\indent
After beam pions are captured, the de-excitation photons will produce electrons and positrons. Those are at the beam-prompt timing, outside the analysis window. In the signal-momentum region, there are 0.09 electrons by muon decay in orbit. There could be backgrounds smaller than 0.027 ($< 0.05\ 90\mathrm{\%\ C.\ L.}$) induced by delayed protons from the accelerator at an irregular timing \cite{4}. But they are much smaller than 1 event per year. \\
\indent
%The analysis window is short enough compared to the cycles of the accelerator $40\ \mathrm{msec}$ so that
Cosmic-ray background from electrons are estimated at $< 0.018$, and from muons $< 0.001$. 
\begin{center}
  \begin{figure}[t]
    \begin{tabular}{ccc}
      \begin{minipage}[t]{0.25\hsize}
        \centering
        \includegraphics[width=4cm,keepaspectratio,clip]{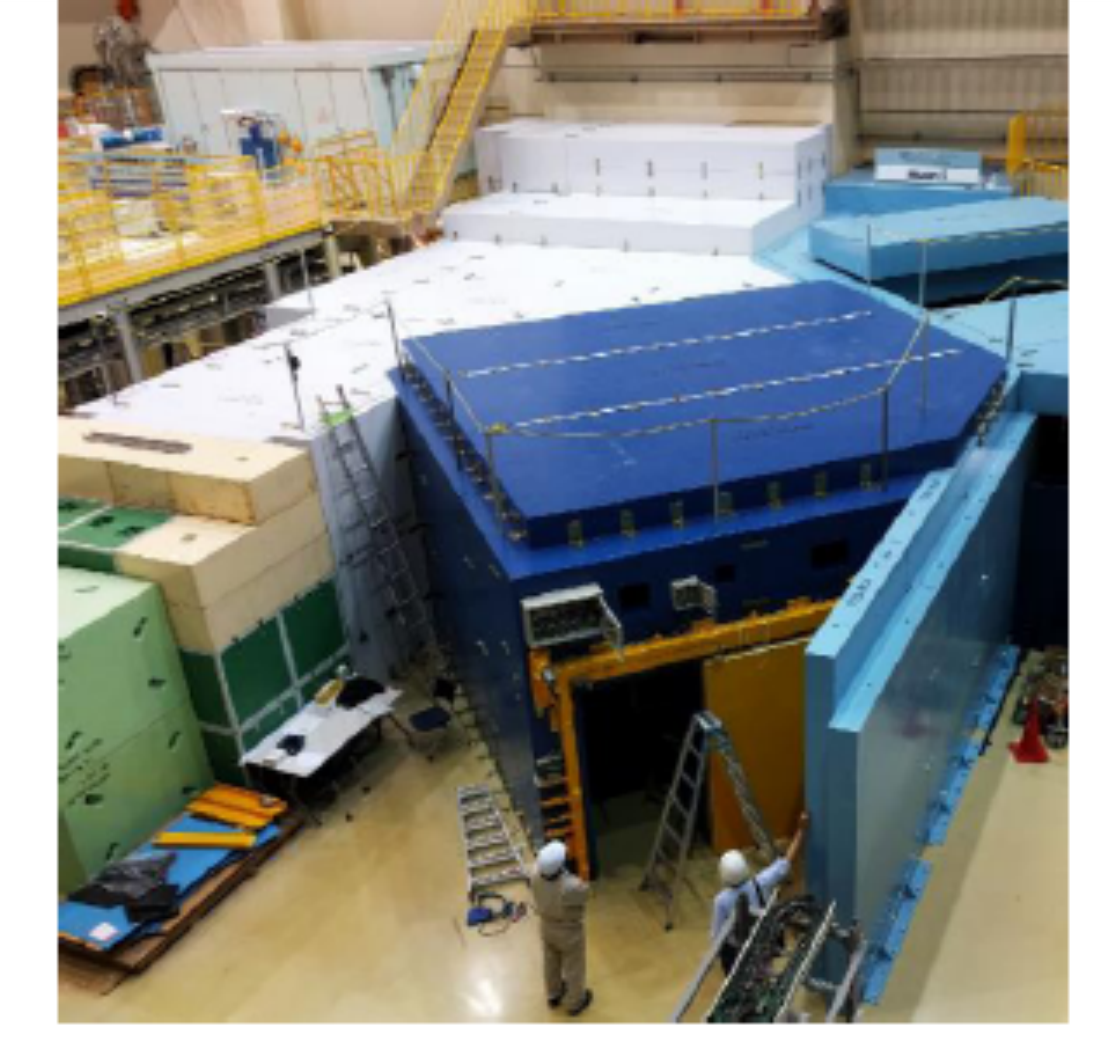}
      \caption{Photograph of H Line (under construction) and H1 Area (deep blue).}
      \label{fig:h1area}
      \end{minipage} &
      \begin{minipage}[t]{0.28\hsize}
        \centering
        \includegraphics[width=3.3cm,keepaspectratio,clip]{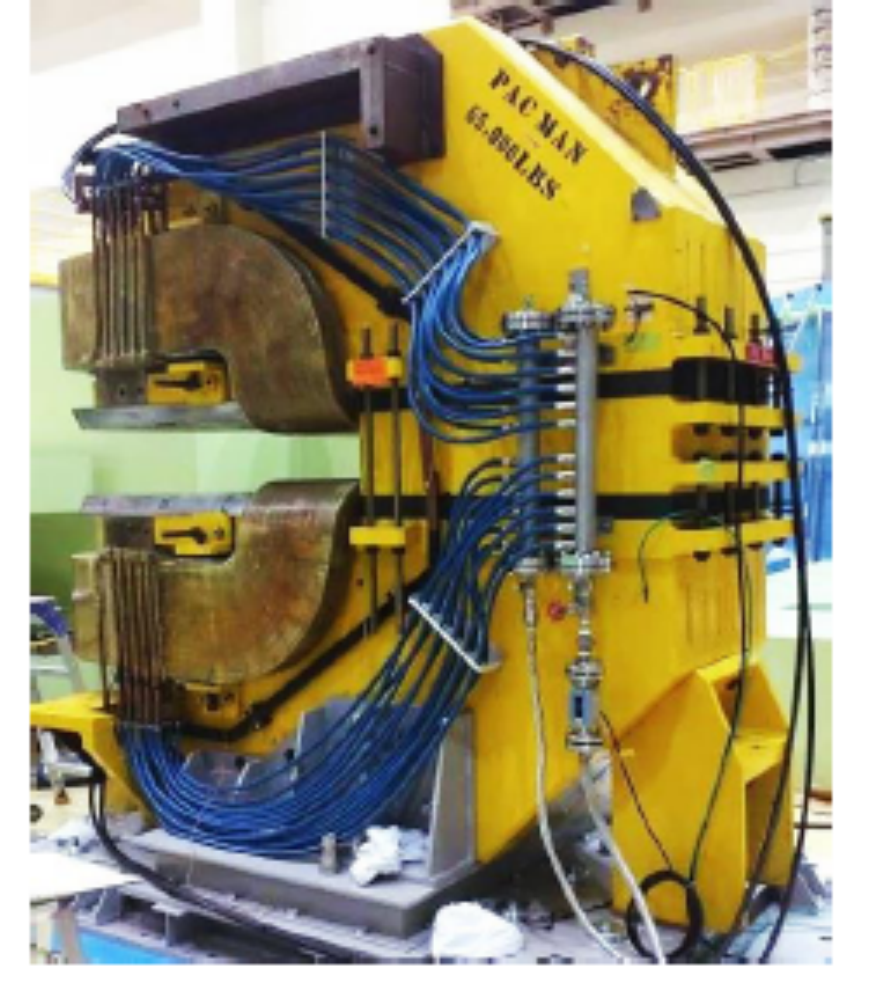}
      \caption{The magnet PACMAN which will be used for the spectrometer.}
      \label{fig:pacman}
      \end{minipage} &
      \begin{minipage}[t]{0.38\hsize}
        \centering
        \includegraphics[width=5.5cm,keepaspectratio,clip]{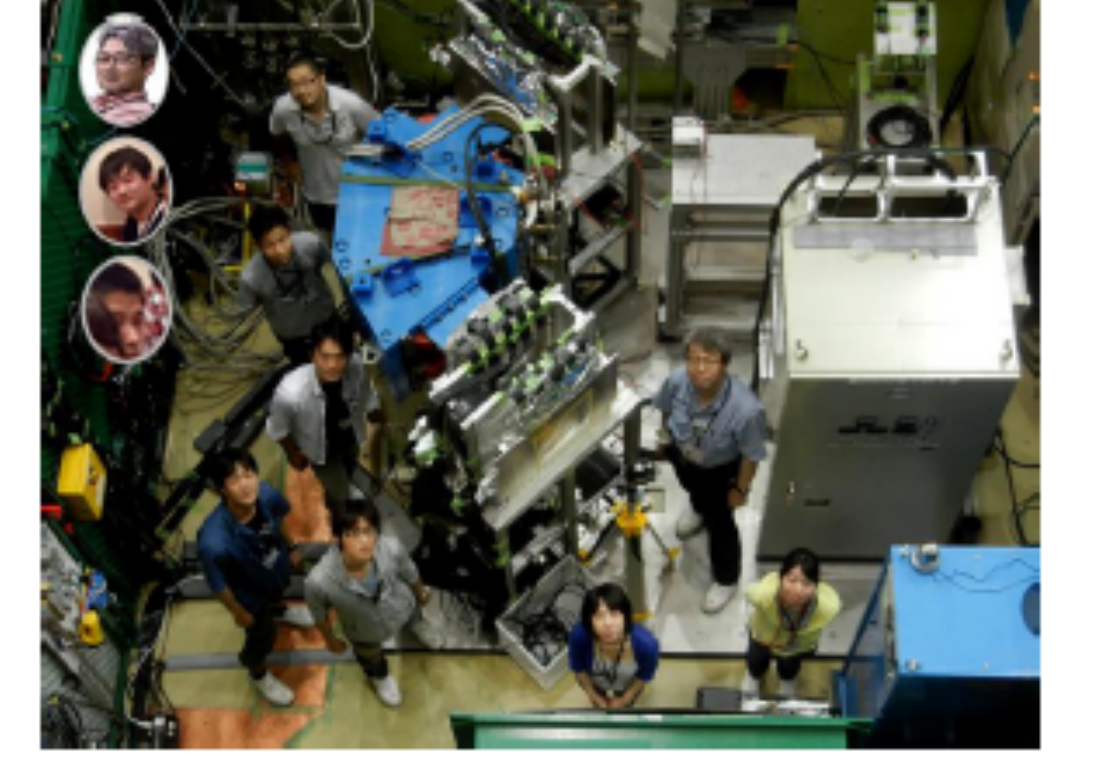}
      \caption{Photograph of experimental setup at the J-PARC MLF D2 Area in 2017.}
      \label{fig:d2}
      \end{minipage}      
    \end{tabular}
  \end{figure}
\end{center}
\vspace{-20pt}
\noindent
There are no backgrounds induced by anti-protons because the proton beam energy of $3\ \mathrm{GeV}$ is below the anti-proton production threshold.
\subsection{Current Status}
The H Line, the secondary beamline which will be used for the DeeMe experiment, is under construction (see Fig. \ref{fig:h1area}). \\
% Magnets of the H Line are being tested on the H1 Area. 
%
%The alignment of magnets is in progress. \\
\indent
For the spectrometer magnet, we will use a dipole magnet that we call PACMAN (Fig. \ref{fig:pacman}). This magnet is leased from TRIUMF. It was tested and worked well. \\
\indent
Tracking detectors, all four of the MWPCs, were manufactured. We use unique 
% high voltage switching MWPC to restore operation quickly after being hit by prompt charged particles and to detect a single electron.
MWPCs with a high-voltage switching mechanism to dynamically control the gas multiplication and to avoid space charge effects due to a large number of prompt charged particles. The MWPCs are quickly restored for the normal operation soon after the prompt timing to detect a signal electron. \\
The hit efficiency was measured to be approximately 98\% for detecting single electrons. More details on the development of the detectors can be found in \cite{5}. \\
\indent
Last year, measurement of momenta about $50\ \mathrm{MeV/}c$ of electrons from muon decay-in-orbit was conducted at the D Line in J-PARC MLF as shown in Fig. \ref{fig:d2}. All four of the MWPCs and DAQ \cite{6} worked well. Analysis codes including waveform handling, hit cluster finding, and track fitting through the spectrometer magnet were developed. The detector system is ready for physics runs to search for the $\mu$-$e$ conversion.
\section{The COMET Experiment}
\label{sec:comet}
%\label{sec:another}
%There is no strict length limitation, but the authors are strongly encouraged to keep contents to the strict minimum necessary for peers to reproduce the research described in the paper.
The COMET Collaboration consists of more than two hundred people representing forty institutes as of November, 2018.
\subsection{The Concept in Phase-I}
\begin{center}
  \begin{figure}[t]
    \begin{tabular}{cc}
      \begin{minipage}[t]{0.45\hsize}
        \centering
        \includegraphics[width=7cm,keepaspectratio,clip]{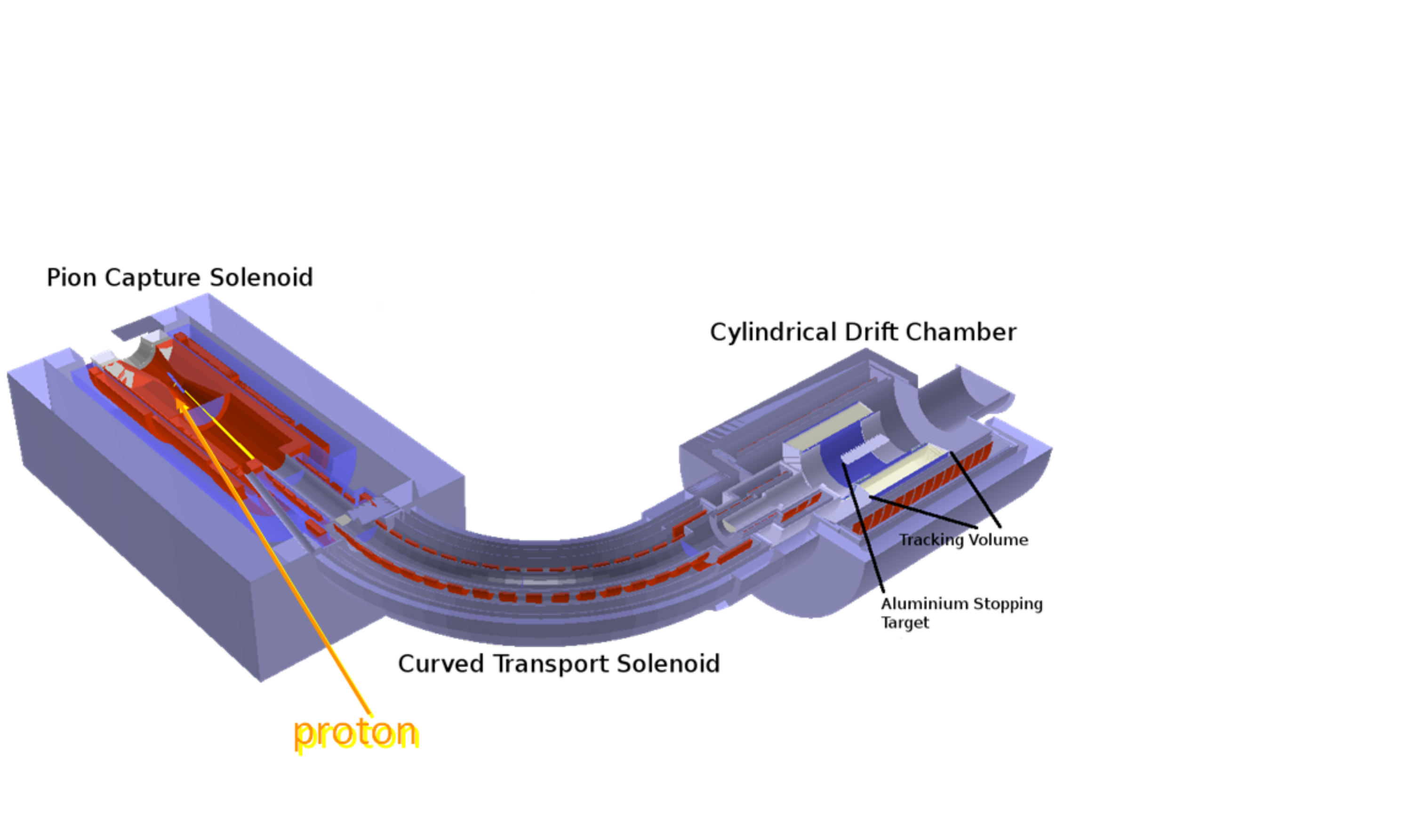}
      \caption{Experimental apparatus of the COMET in Phase-I at the Hadron Hall.}
      \label{fig:cometphase1}
      \end{minipage} &
      \begin{minipage}[t]{0.45\hsize}
        \centering
        \includegraphics[width=6cm,keepaspectratio,clip]{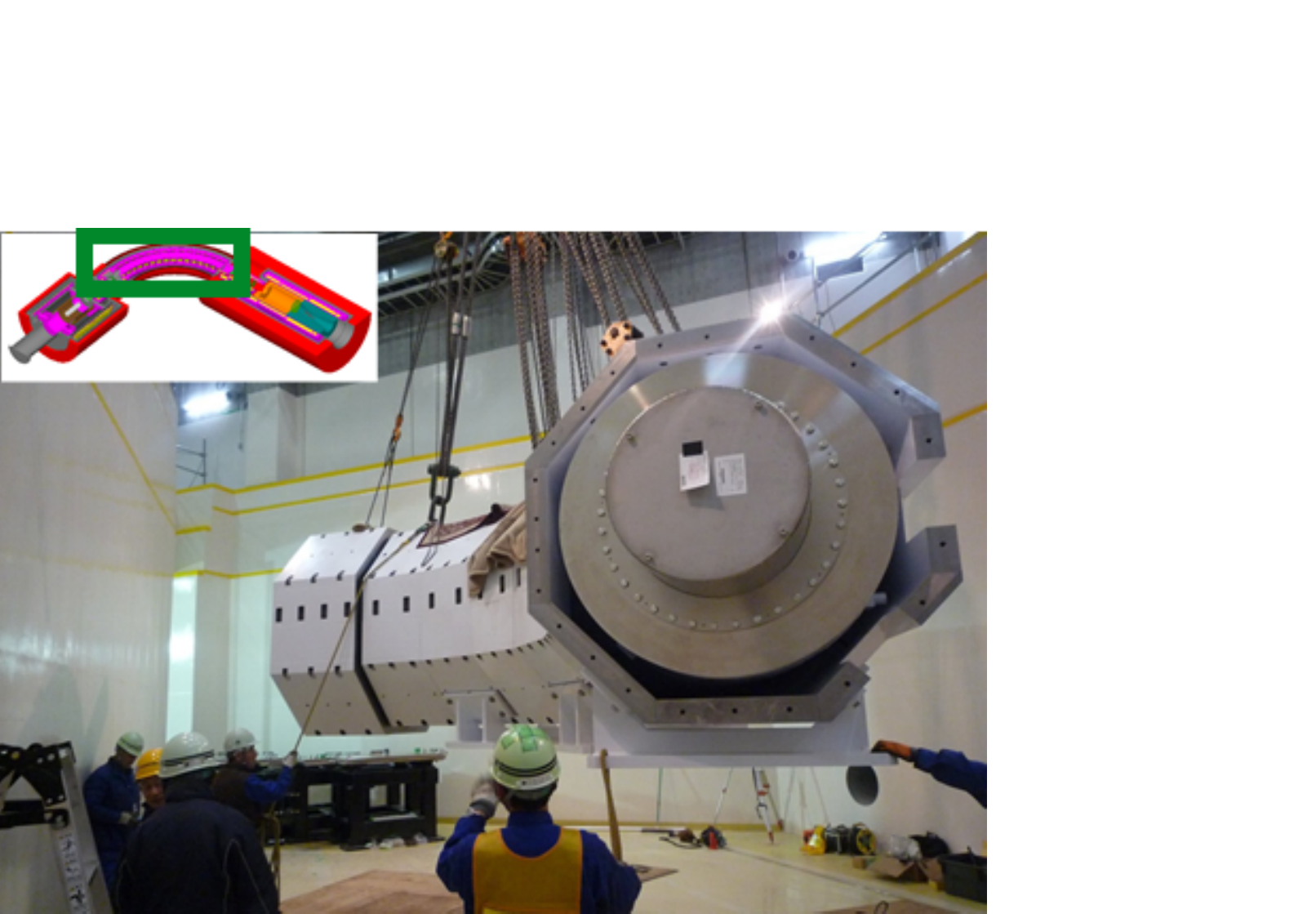}
      \caption{Installed muon transport solenoid.}
      \label{fig:muontransportsolenoid}
      \end{minipage}
    \end{tabular}
  \end{figure}
\end{center}

Figure \ref{fig:cometphase1} illustrates the experimental apparatus of the COMET in Phase-I. The 8-GeV, 3.2-kW pulsed proton beam from the Main Ring will be used to produce pions, which will decay to muons. The muon transport solenoid then transports muons by 90-degree bending to select low-momentum ones. \\
\indent
The Cylindrical Drift Chamber (CDC) is used as the main detector. The requirement is to have a momentum resolution better than $200\ \mathrm{keV/}c$. For the background measurement, the straw tube tracker and electron calorimeter (StrawECAL) will be used \cite{7}.
\vspace{-20pt}
\subsection{Status and Recent Highlights}
The experiment will be conducted in a new south building at the Hadron Experimental Facility. The experimental facility was buit in 2015. \\
\indent
The Muon Transport Solenoid (Fig. \ref{fig:muontransportsolenoid}) was installed at the Hadron Hall. It is ready for cryogenic tests. StrawECAL, COMET-CDC and its prototype trigger system are being tested using beam or cosmic-rays as shown in Fig. \ref{fig:cometcdc}. Those detectors are under construction and development, and trying to be ready to start in 2019. \\
\indent
Recently 
% it was tested to slow-extract 8-GeV pulsed protons to the Hadron Hall, and to understand beam-related backgrounds.
the slow extraction of 8-GeV pulsed protons to the Hadron Hall was tested and beam-related backgrounds were investigated. The extinction of protons between two main pulses was found to be good enough for the experiment.
\subsection{Toward Phase-II}
\begin{center}
  \begin{figure}[t]
    \begin{tabular}{cc}
      \begin{minipage}[t]{0.45\hsize}
        \centering
        \includegraphics[width=5.5cm,keepaspectratio,clip]{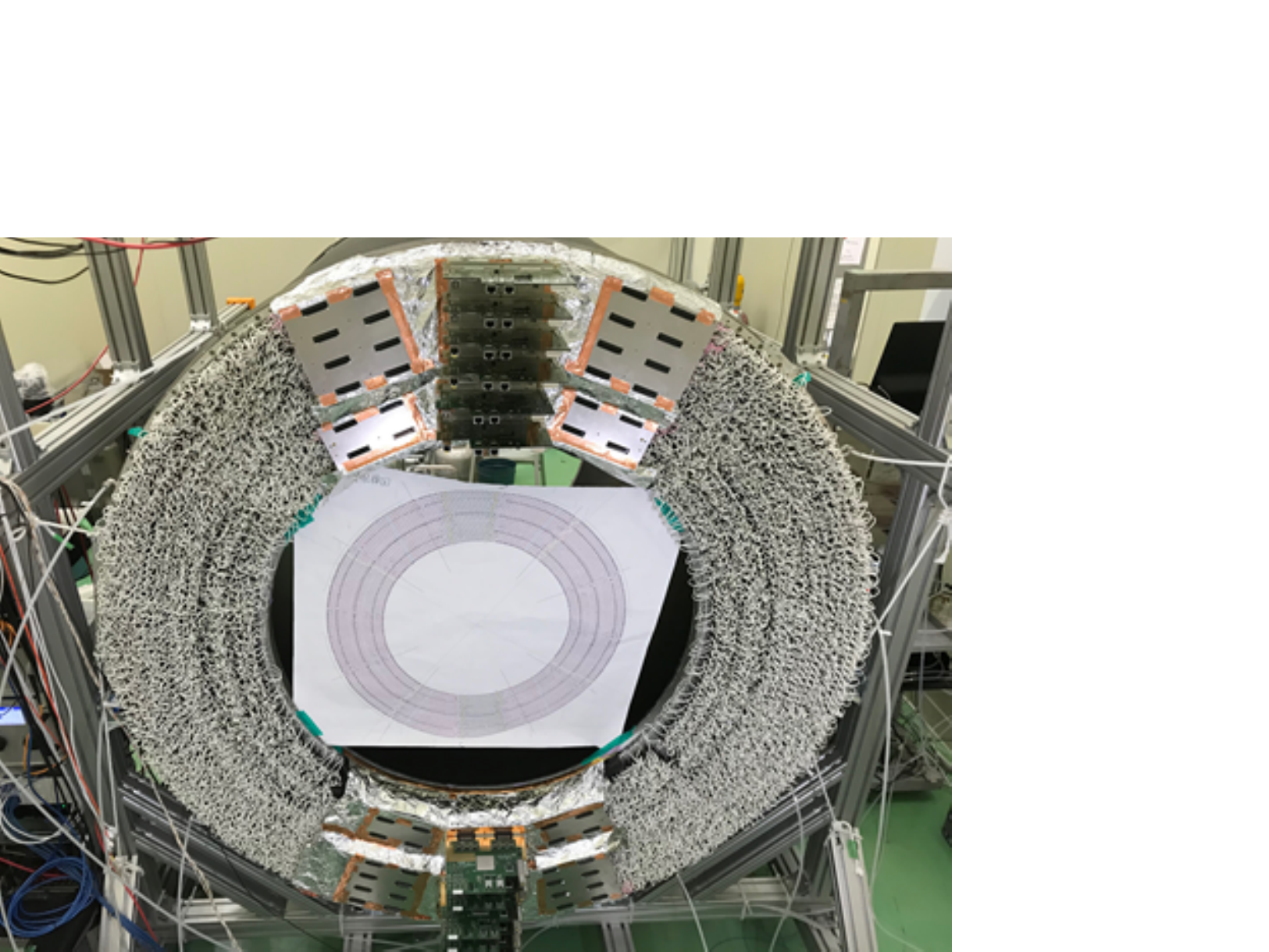}
        \caption{Photograph of the Cylindrical Drift Chamber (CDC) and its trigger system (prototype) for COMET.}
        \label{fig:cometcdc}
      \end{minipage} &
      \begin{minipage}[t]{0.45\hsize}
        \centering
        \includegraphics[width=7.3cm,keepaspectratio,clip]{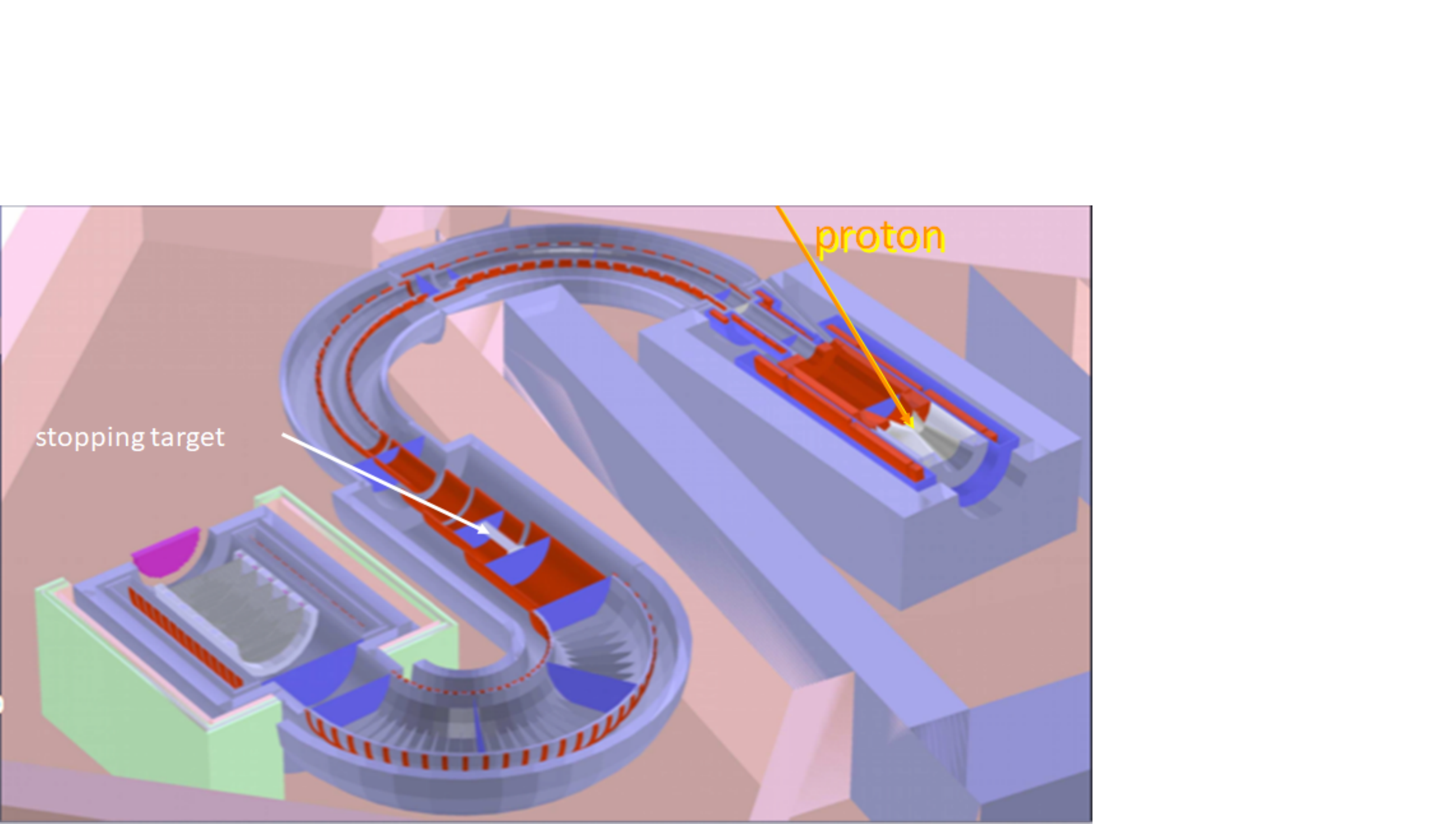}
        \caption{Layout of the COMET experiment in Phase-II.}
        \label{fig:cometphase2}
      \end{minipage}
    \end{tabular}
  \end{figure}
\end{center}

The Phase-II study is also in progress. Changing the transport solenoid design to S-shape, using a C-shaped muon transport and a C-shaped electron spectrometer (Fig. \ref{fig:cometphase2}), and increasing the proton beam intensity to $56\ \mathrm{kW}$, COMET will aim further improvement of sensitivity by two orders of magnitude from the experiment in Phase-I.
\section{Conclusion}
There are two plans of experiments to search for the $\mu$-$e$ conversion at J-PARC, which are called DeeMe and COMET. The DeeMe aims at a single event sensitivity of $1\times 10^{-13}$ using a primary production target made of graphite as a muon-stopping target for one year. If we change the target to silicon carbide, the sensitivity will be $2\times10^{-14}$. The spectrometer consisting of a magnet and four MWPCs, is ready. Construction of the secondary beamline, H Line, is in progress at J-PARC Materials and Life Science Experimental Facility (MLF). DeeMe is planning to start data taking soon after the completion of the H Line construction, hopefully in 2019. The COMET will be conducted with a single event sensitivity of $3\times10^{-15}$ in Phase-I or $2\times10^{-17}$ in Phase-II. Preparation of the beam line and detectors is ongoing. The detectors for the experiment in Phase-I will be ready in 2019.
\vspace{-20pt}
\section*{Acknowledgements}
I am very grateful to the COMET Collaboration, especially to Dr. H. Yoshida, Osaka University for giving me information on the COMET experiment.

\nolinenumbers

\end{document}